\documentstyle[prl,aps,epsfig]{revtex}
\begin{document}
\draft
%
%%%%%%%%%%
%
%     Title Page
\year 1998\month 9\day 18
%
%%%%%%%%%%
%
\title{ Optical Spin Vortex in Nonlinear Anisotropic Media}
\author{Hiroshi Kuratsuji and Shouhei Kakigi}
\address{Department of Physics, Ritsumeikan University-BKC, 
               Kusatsu City 525-8577, Japan}
\date{\today}
\maketitle
\begin{abstract} 
A theoretical study is given of a new type of optical vortex in 
nonlinear anisotropic media. This is called an ``optical spin vortex'', 
which is realized as a special solution of the two-component non-linear 
Schr\"{o}dinger equation. The existence of spin vortex is inherent in 
the spin texture that is caused by the anisotropy of dielectric tensor, 
where a role of spin is played by the Stokes vector. 
By introducing the effective Lagrangian governing the non-linear 
Schr\"{o}dinger equation, we derive the evolution equation for 
the motion of spin vortex, which exhibits a clear distinction from the 
conventional singular vortex inherent in the phase function that is 
described by the single component complex field. 
\end{abstract}

%\bigskip 
%\paragraph*{Introduction}
The nonlinear optics has been a major subject in physics for long 
time \cite{Chiao}. The main interest is focussed on particular 
solutions of the basic non-linear equation that governs the light 
field or eletro-magnetic field. Among others, remarkable is the 
optical vortex, the existence of which has been early suggested in 
\cite{Chiao}. Recently the detailed study has been carried out 
from both of theoretical and experimental point of view 
\cite{Snyder,Swartz}. More recently experimental verification 
has been also given for the multi-vortices \cite{Rozas}. 
The basic idea of the optical vortex follows an analogy with the 
existence of the superfluid vortex that is described by the complex 
Landau-Ginzburg order parameter for bose fluid as is stated in 
\cite{Swartz}. The equation for the light field, which is known as 
the nonlinear Schr\"{o}dinger (NLS) equation, is very similar to 
the Pitaevki equation. Thus it is natural to expect the occurrence of 
the optical counterpart of the superfluid vortex. 

The purpose of this letter is to put forward a possible new type of 
vortex and its dynamical behavior in the nonlinear and anisotropic media. 
We call this new type vortex the ``optical spin vortex''. Our starting 
equation is the two-component NLS equation. Here the basic ingredient 
is the state of polarization, which is realized by the two-component 
field. The concept of polarization is fruitful which forms a basis of 
modern crystal optics \cite{Born,Landau}. The quantity that describes 
polarization state is given by the Stokes parameters (or vector), 
which is geometrically realized as a point on the Poincar\'{e} sphere. 
The crucial point is that the Stokes vectors form the field of 
pseudo-spin, namely, we consider the special configuration of 
pseudo-spin. Having recognized the role of the Stokes vector as 
a pseudo-spin, we have the analogy with the spin vortex. This can be 
naturally formulated by the effective Lagrangian for the pseudo-spin 
field reduced from two-component NLS equation. Such a Lagrangian has 
not been used thus far in condensed matter physics. Notable here 
is that the spin vortex is non-singular in contrast to the singular 
vortex that appears in the conventional optical vortex, where the 
singularity is inherent in the phase of the complex scalar field. 
By using the effective Lagrangian, we can naturally derive the 
evolution equation for the non-singular vortex, which reveals the 
topological invariant that is characteristics of the non-singular 
vortex. In this note, we are concerned with the basic qualitative 
idea and the details of numerical analysis will be given elsewhere.

%\bigskip
\paragraph*{Two component nonlinear Schr\"{o}dinger equation.---}
First we shall derive the two-component nonlinear Schr\"{o}dinger 
equation for the light wave traveling through anisotropic media. 
The procedure follows the one developed 
in a recent paper \cite{Kakigi}. Suppose that 
the electro-magnetic wave of wave vector $k$ travels in the direction 
of $z$ with the dielectric tensor $\hat \epsilon$. The nonlinear 
nature of media implies that $\hat \epsilon$ has a field dependence 
in nonlinear form. The $z$-axis is prescribed to be the principal 
axis of the dielectric tensor, namely, the axis corresponding to 
one of the eigenvalues of the dielectric tensor. 
In this geometry, $\hat \epsilon$ is taken to be $2 \times 2$ matrix 
and the Maxwell equation for the displacement field ${\bf D}$ is given 
by the second order differential equation: 
\begin{equation}
{\partial^2 {\bf D} \over \partial z^2} + \nabla^2{\bf D} 
+ \omega^2\hat\epsilon{\bf D} = 0 
\label{one}
\end{equation}
where $\nabla = ({\partial\over\partial x},\; {\partial\over\partial y})$
and $(x,y)$ denotes the coordinate in the plane perpendicular to $z$
axis. Now we put
\begin{equation}
{\bf D}(z) = {\bf f}(x, y, z)\exp[ikn_0z]
\label{two}
\end{equation}
with $k = {\omega\over c}$, and $n_0 (\equiv\sqrt{\epsilon_0})$ 
means the refractive index for the otherwise isotropic medium. 
The amplitude ${\bf f}(x, y, z)$ is written as 
${\bf f} = {}^t(f_1, f_2) = f_1 {\bf e}_1 + f_2{\bf e}_2$, 
which is slowly varying function of $z$ besides the $(x, y)$ 
dependence, and ${\bf e}_1$ and ${\bf e}_2$ denotes the basis of 
linear polarization. By subsituting (\ref{two}) into (\ref{one}) 
and considering the short wave length limit, i.e.,
$\left\vert{\partial{\bf f} \over \partial z}\right\vert << k$, 
we can derive the equation for the amplitude ${\bf f}$; 
namely, keeping only the first derivative ${\partial{\bf f} \over 
\partial z}$ besides the Laplacian with respect to $(x,y)$, 
we get 
\begin{equation}
i\lambda{\partial {\bf f} \over \partial z} + 
\left[{\lambda^2 \over n_0^2}\nabla^2 + (\hat {\epsilon} - 
n_0^2)\right]{\bf f} = 0
\end{equation}
where $\lambda$ is the wavelength divided by $2\pi$. 
This equation is regarded as a two-state Schr\"{o}dinger equation 
where $\lambda$ just plays a role of the Planck constant and $z$ is 
a substitute of the time variable. The components $(f_1, f_2)$ couple 
each other to give rise to the change of polarization which is just 
the effect of birefringence governed by a $2\times 2$ matrix 
``potential'' $\hat v = \hat\epsilon - n_0^2$. 
The $\hat v$ represents the deviation from the isotropic value 
and if the non-abosortive media is considered, it becomes 
hermitian.  From the hermiticity, the most general form of 
$\hat v$ is written as 
\begin{equation}
{\hat v} = \left(
 \begin{array}{cc}
  v_0 + \alpha & \beta + i\gamma \\
  \beta -i\gamma & v_0 -\alpha \end{array}
\right) .
\end{equation}
Now for the use of later argument, it is convenient to transform the basis 
of linear polarization ${\bf f}$ in the circular basis, that is, 
\begin{equation}
{\bf e_{\pm}} = {1 \over \sqrt 2}({\bf e}_1 \pm i{\bf e}_2)
\end{equation}
which defines the unitary transformation $({\bf e}_{+}, {\bf e}_{-})
= T({\bf e}_1, {\bf e}_2)$ Here $T$ is the unitary 
transformation of $2 \times 2$ matrix: 
\begin{equation}
 T = {1 \over \sqrt 2} \left(
 \begin{array}{cc}
 1 & i \\
 1 & -i \end{array}
\right) .
\end{equation}
Hence we can introduce the wave function and as 
$\psi = T{\bf f}= {}^t(\psi_1^{*}, \psi_2^{*})$. 
Thus we have 
\begin{equation}
i\lambda{\partial \psi \over \partial z} = \hat H \psi 
\end{equation}
and the transformed ``Hamiltonian'' becomes 
\begin{equation}
\hat H = T h T^{-1} = -{\lambda^2 \over n_0}\nabla^2 + V
\end{equation} 
where the potential term is written in terms of the Pauli spin; 
$V = v_0\times 1 + \sum_{i=1}^3v_i\sigma_i$. 
We here introduce the ``quantum'' Lagrangian leading to the 
Schr\"{o}dinger type equation, which is given by 
\begin{equation}
 I = \int \psi^{\dagger}\left(i\lambda{\partial \over \partial z} - 
\hat H\right)\psi d^2x dz .
\end{equation}
Indeed, the Dirac variation equation $\delta I = 0$ recovers 
the Scr\"odinger equation. For later convenience, 
we write $L_C = \int \psi^{\dagger}i\lambda{\partial \over \partial z}
\psi d^2x$ and $H = \int \psi^{\dagger}\hat H\psi d^2x$, which are 
called the canonical term and the Hamiltonian term respectively.

%\bigskip
\paragraph*{Reduction to the pseudo-spin field.---} 
Now, having given the two component field function $\psi$, 
we can introduce the Stokes parameter by using the Pauli 
spin \cite{Brosseau}, namely, this is written as 
$S_i = \psi^{\dagger}\sigma_i\psi, 
S_0 = \psi^{\dagger}1\psi$ with $i = x,y,z$ . 
We see that the relation $S_0^2 = S_x^2 + S_y^2 + S_z^2$ 
holds. $S_0$ gives the field strength; $S_0^2 \equiv 
\vert{\bf D}\vert^2$. Using the spinor representation,
\begin{equation}
\psi_1 =S_0 \cos{\theta \over 2},\;
\psi_2 =S_0 \sin{\theta \over 2}\exp[i\phi] ,
\end{equation}
we have the polar form for the Stokes vector ${\bf S} = (S_x, S_y, S_z)$ ;
\begin{equation}
S_x =S_0 \sin\theta \cos\phi,\; S_y = S_0\sin\theta \sin\phi,\; 
S_z =S_0 \cos\theta 
\end{equation}
which forms a pseudo spin, so to speak, and is pictorially given by 
the point on the Poincar\'{e} sphere. In terms of the angle variable, 
the canonical and Hamiltonian term in the Lagrangian are given as 
\begin{eqnarray}
L_C & = & \int {S_0^2\lambda \over 2}(1 - \cos\theta)
{\partial \phi \over \partial z}d^2x , \nonumber\\
  H & = & H_T + \tilde V .
\label{three}
\end{eqnarray}
Here the potential term in the Hamiltonian is given as 
\begin{equation}
\tilde V = \int (v_0 + \sum_{i=1}^{3}v_iS_i)d^2x  .
\end{equation}
We note that $v_0$ and $v_{i}$'s are nonlinear functions of the field 
strength $S_0$ in general and this fact suggests that we have a stable 
special solution for the field of pseudo spin. The kinetic energy term is 
written as a sum of three terms; 
$H_T = {\lambda^2 \over n_0 }\int \nabla \psi^{\dagger}\nabla \psi d^2x
\equiv H_1 + H_2 + H_3$ where the respective terms become  
\begin{eqnarray}
H_1 & = & \int {S_0^2\lambda^2 \over n_0}(\nabla S_0)^2d^2x , \nonumber\\
H_2 & = & \int {S_0^2\lambda^2 \over 4n_0}\{(\nabla\theta)^2 
+ \sin^2\theta(\nabla\phi)^2\}d^2x , \nonumber\\
H_3 & = \int & {S_0^2\lambda^2 \over4 n_0}
\{(1-\cos\theta)\nabla\phi\}^2d^2x 
\label{pseudo}
\end{eqnarray}
where $H_1$ gives the kinetic energy associating with the space 
modulation of the field strength, and the second term represents 
the intrinsic energy for pseudo-spin which exactly coincides with 
the continuous Heisenberg spin chain \cite{Ono}. The last term is 
nothing but the fluid kinetic energy inherent spin structure, 
where we put 
\begin{equation}
 {\bf v} = (1 -\cos\theta)\nabla\phi 
\label{velocity}
\end{equation}
which means the velocity field as is seen below. In this way, we have 
a new type of fluid, so to speak, ``anisotropic optical fluid'', which 
is naturally described by the Lagrangian written in terms of the 
pseudo-spin field. 

\begin{figure}[htpb]
\begin{minipage}[b]{8.2cm}
\epsfig{file=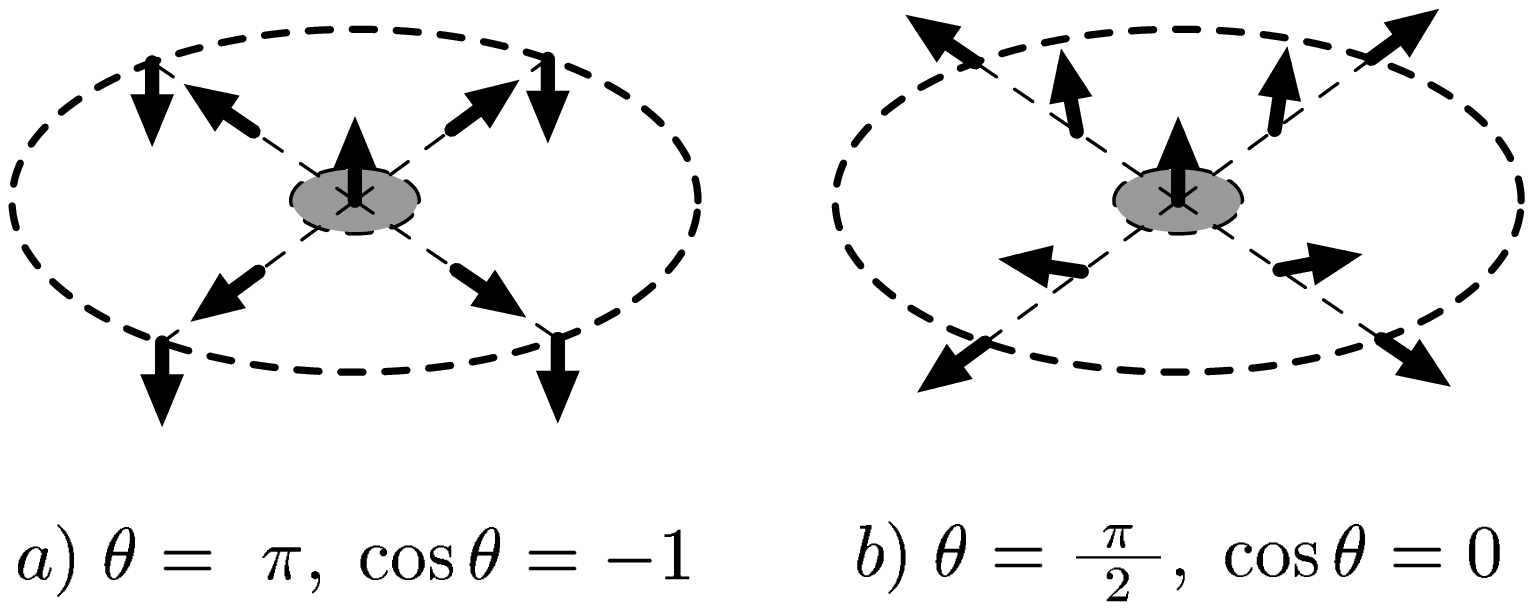,width=8.2cm}
\caption{Profile of the non-singular vortex; a) The case of 
$\theta(\infty) = \pi$ and b) The case of $\theta(\infty) = {\pi \over 2}$.}
\label{fig1}
\end{minipage}\hfill
\begin{minipage}[b]{8.2cm}
\epsfig{file=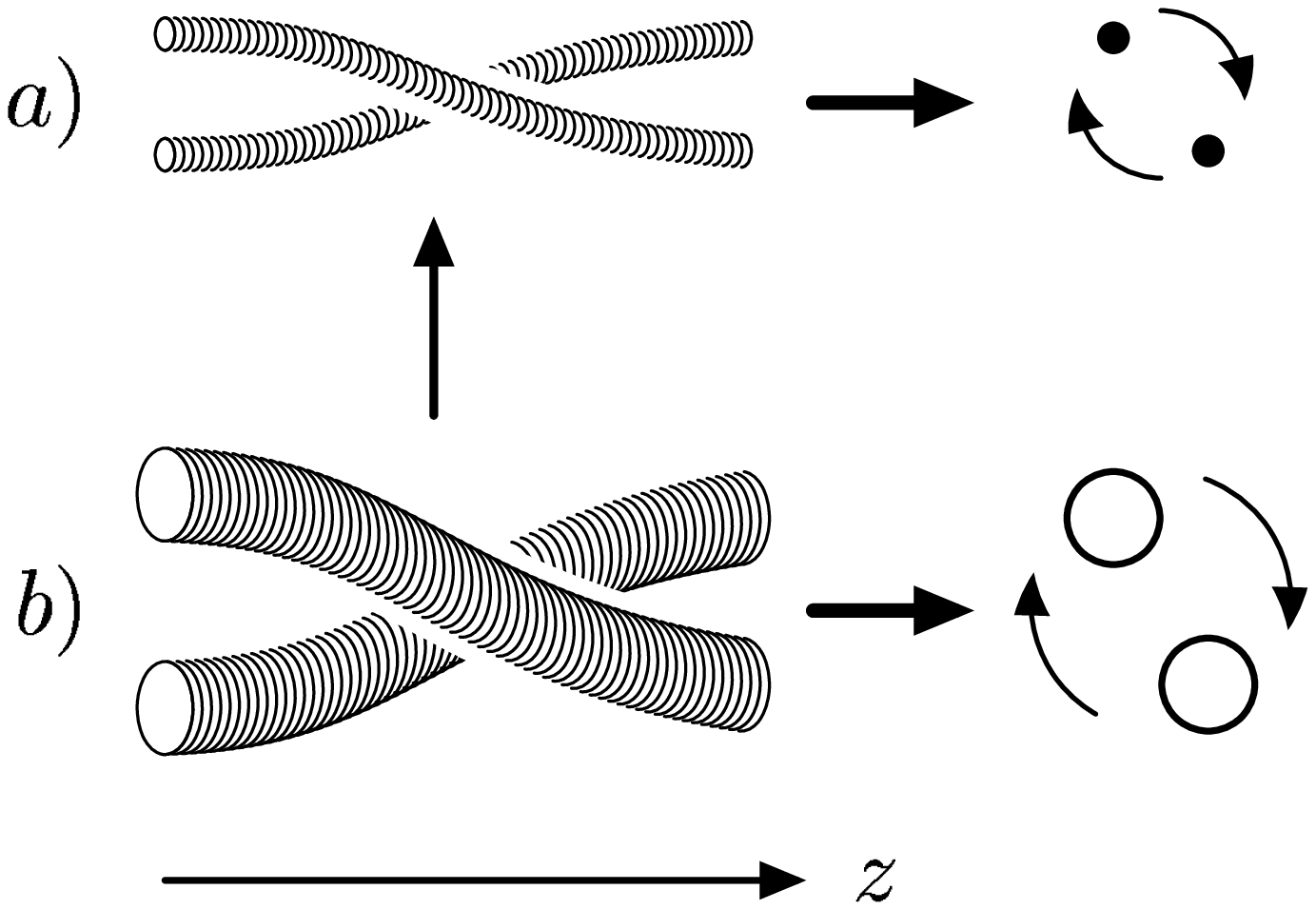,width=8.2cm}
\caption{Illustration of for evolution of vortex pairs and the 
possible transition between the non-singular (b) and the singular (a) 
vortex. The bullet $\bullet$ and white circle $\bigcirc$ in the right 
hand side indicate an instantaneous vortex pair at a particular value 
of $z$.}
\label{fig2}
\end{minipage}
\end{figure}

%\bigskip
\paragraph*{Equation of motion for optical spin vortex.---} 
We now examine the vortex solution by starting with the pseudo-spin 
Lagrangian given in the above. Taking variation leads to the coupled 
equations for the field variables $(S_0, \theta, \phi)$. In what follows, 
we confine our argument to the case that $S_0$ becomes constant. 
Physically, this corresponds to the constant background field with 
a ``dark'' core which is controlled by the profile of the remaining 
variables $(\theta, \phi)$. A static solution for one vortex is obtained 
with the phase function $\phi = \tan^{-1}{y \over x}$, and the profile 
function $\theta$ is given as a function of the radial variable $r$. 
The explicit form of the profile function $\theta(r)$ may be derived 
from the extremum condition for the Hamiltonian $H$, with the specific 
boundary condition at $r=\infty$ and $r=0$. Namely, we choose such that 
at the origin $r=0$, we can take $\theta(0)=0$ in general, that is, the 
left-handed circular polarization at the origin [see Fig.\ref{fig1}(a) 
and (b)], whereas at $r=\infty$, there exist several possibilities; 
we here consider two typical cases: 
A) $\theta(\infty)= \pi$ and B) $\theta(\infty) =\pi/2$. 
We here introduce the vector ${\bf m}(x) \equiv {\bf S}/S_0$, 
Then, we have $m_3(0)=1$ for both cases, A), B), namely, the spin field 
directs upward. On the other hand, we have at $r=\infty$, 
$m_3(\infty)=-1$ for case A) and $m_3(\infty)=0$ for case B), that is, 
the spin field directs {\it downward} for A), which represents 
the right-handed circular polarization [Fig.\ref{fig1}(a)] and 
{\it outward} for B), which represents the linear polarization 
[Fig.\ref{fig1}(b)]. Here we simply assume the existence of the solution 
$\theta(r)$satisfying the above boundary condition. 
%%%%%%%%%%%%%%%%%%%%%%%%%%%%%%%%%%%%%%%%%%%%%%%%%%%%%%%%%%%%%%%%%%%%%%
%%                                                                  %%
%%   Put Fig. 1 around here                                         %%
%%                                                                  %%
%%%%%%%%%%%%%%%%%%%%%%%%%%%%%%%%%%%%%%%%%%%%%%%%%%%%%%%%%%%%%%%%%%%%%%
%% \begin{figure}                                                   %%
%% \caption{Profile of the non-singular vortex; a) The case of      %%
%% $\theta(\infty) = \pi$ and b) The case of $\theta(\infty) = {\pi %%
%% \over 2}$.}                                                      %%
%% \label{fig1}                                                     %%
%% \end{figure}                                                     %%
%%%%%%%%%%%%%%%%%%%%%%%%%%%%%%%%%%%%%%%%%%%%%%%%%%%%%%%%%%%%%%%%%%%%%%

In order to treat the motion for a single vortex, we introduce the 
coordinate of the center of vortex, ${\bf R}(z) = (X(z), Y(z))$, 
by which the vortex solution is parameterized such that 
$\theta({\bf x} - {\bf R}(z))$ and $\phi({\bf x} - {\bf R}(z))$. 
By using the parameterization prescribed in the above, the canonical 
term $L_C$, the first term in (\ref{three}), is written as 
\begin{eqnarray}
 L_C & = & \int{S_0^2\lambda \over 2}(1 - \cos\theta)
\nabla\phi\cdot\dot{\bf R}d^2x \nonumber\\ 
   & = & {S_0^2\lambda \over 2}\int {\bf v}\cdot \dot{\bf R}d^2x
\label{seven}
\end{eqnarray}
where we have used the relation: ${\partial \phi \over \partial z} 
= {\partial \phi \over \partial {\bf R}}\dot{{\bf R}}, 
{\partial \phi \over \partial{\bf R}} = -\nabla\phi$ 
with $\dot {\bf R} \equiv {d{\bf R} \over dz}$. Remarkable point here 
is that the velocity field (\ref{velocity}) does not bear singularity 
at the origin compared with the behavior of the conventional vortex 
for which the velocity field is simply defined by the gradient of 
the phase ${\bf v} = \nabla \phi$ \cite{Swartz}. 
Keeping this in mind, we shall derive the equation of motion for 
a single vortex. In order to achieve this we consider the 
Euler-Lagrange equation for ${\bf R}$, which gives the ``balance of forces'' 
\begin{equation}
{\bf F}_C \equiv 
{d \over dz}{\partial L_C \over \partial \dot{\bf R}} - 
      {\partial L_C \over \partial {\bf R}} 
       = - {\partial H \over \partial {\bf R}} .
\label{nine}
\end{equation}
By using eq.(\ref{velocity}), we get 
\begin{equation}
{S_0^2\lambda \over 2}\sigma ({\bf k} 
\times \dot{\bf R}) = - {\partial H \over \partial {\bf R}}
\label{nineA}
\end{equation}
where the ${\bf k }$ in (\ref{nineA}) is the unit vector perpendicular 
to the $xy$-plane. Here $\sigma$ is defined as 
\begin{equation}
\sigma = \int\left({\partial v_y \over \partial x} 
       - {\partial v_x \over \partial y}\right)d^2x .
\label{vorticity}
\end{equation}
In deriving (\ref{nineA}), we have used the relation 
${\partial v_x \over \partial X} = -{\partial v_x \over \partial x}$. 
The integrand of $\sigma$ is nothing but the vorticity which 
we put $\omega$. Using eq.(\ref{velocity}), we can write $\omega$ 
in terms of the angular variable: 
\begin{equation}
 \omega = \nabla \times {\bf v} = 
 \sin\theta(\nabla\theta \times \nabla\phi) 
\label{twel}
\end{equation} 
which is also rewritten in terms of the spin field ${\bf m}$
\begin{equation}
\nabla \times {\bf v} = {\bf m}\cdot\left({\partial{\bf m} \over 
\partial x}\times{\partial {\bf m} \over \partial y}\right) .
\label{thirtA}
\end{equation}
Equation (\ref{thirtA}) (or (\ref{twel})) is an optical counterpart of 
a topological invariant of hydrodynamical origin \cite{Lamb} or otherwise 
the Mermin-Ho relation for superfluid He3-A \cite{Volovik}. 
We get an integral of the differential form; 
\begin{equation} 
\sigma = \int_S \sin\theta d\theta\wedge d\phi .
\label{thirt}
\end{equation}
Here taking into account of the boundary conditions for the profile 
function $\theta(r)$ mentioned above, we have a topological 
interpretation for $\sigma$. In the case A), the vortex configuration 
gives the continuous mapping from the compactified space $S_2$ to 
the spin configuration $S_2$, so the $\sigma$ in (\ref{thirt}) has 
a meaning of the mapping degree of $S_2 \rightarrow S_2$. Hence we get 
the quantization of $\sigma$: $\sigma=m$ ($m$=integer). For the case B), 
on the other hand, the mapping becomes 
$S_2 \rightarrow S_2/2 ({\rm hemisphere})$, so intuitively we have 
the quantization $\sigma=m/2$ ($m$: integer). Thus it should be noted 
that this distinguishes the topological invariant for the conventional 
optical vortex \cite{Swartz}, which gives the invariant of 
$\oint_C \nabla\phi d{\bf s} = 2\pi$. It should be noted that the left 
hand side of the equation of motion (\ref{nineA}) is considered to be 
an optical analog of the Magnus force in superfluids or superconductors. 
As is mentioned above, the ``force'' incorporates the topological 
structure inherent in the coreless vortex, which is quite different 
from the conventional optical vortex which is singular at the origin. 
The point here is that there are two distinct types vortices according to 
the different boundary conditions at infinity. 

The difference between two types of vortex is significant from the point 
of view of condensed matter physics, namely, we expect that the 
non-singular spin vortex is converted to the usual phase vortex. 
This may be considered to be a kind of phase transition as a consequence 
of the phase transition from anisotropic and isotropic that occurs in 
the medium. Actually, as a result of transition the birefringence 
disappears and the Hamiltonian (or dielectric tensor) becomes scalar 
matrix, which means that the two component equation is converted to 
one component; $\psi = S_0\exp[i\phi]$, where the amplitude vanishes at 
the vortex center. In the process of phase transition the core structure 
of the optical vortex may survive and it may be possible to observe 
the structure change between two cores; the soft and singular [See 
Fig.\ref{fig2}]. 
%%%%%%%%%%%%%%%%%%%%%%%%%%%%%%%%%%%%%%%%%%%%%%%%%%%%%%%%%%%%%%%%%%%%%%
%%                                                                  %%
%%   Put Fig. 2 around here                                         %%
%%                                                                  %%
%%%%%%%%%%%%%%%%%%%%%%%%%%%%%%%%%%%%%%%%%%%%%%%%%%%%%%%%%%%%%%%%%%%%%%
%% \begin{figure}                                                   %%
%% \caption{Illustration of for evolution of vortex pairs and the   %%
%% possible transition between the non-singular (b) and             %%
%% the singular (a) vortex. The bullet $\bullet$ and white circle   %%
%% $\bigcirc$ in the right hand side indicate an instantaneous      %%
%% vortex pair at a particular value of $z$.}                       %%
%% \label{fig2}                                                     %%
%% \end{figure}                                                     %%
%%%%%%%%%%%%%%%%%%%%%%%%%%%%%%%%%%%%%%%%%%%%%%%%%%%%%%%%%%%%%%%%%%%%%%

%\bigskip
\paragraph*{The case of multi-vortices.---} 
We now consider the correlated motion of two or more vortices. 
The canonical term in Lagrangian is simply obtained by replacement as 
${\bf v}\cdot \dot{\bf R} \rightarrow \sum_i {\bf v}_i\cdot \dot{\bf R_i}$ 
where ${\bf v}_i$ is the velocity field coming from the i-th vortex. 
The effective Hamiltonian for the assembly of vortices is given by 
the fluid kinetic energy, \cite{potential} namely, 
\begin{eqnarray}
H_{eff} & = & \int \sum_{ij}\omega({\bf y'})
\nabla\log\vert {\bf y'} -({\bf x}- {\bf R}_i)\vert 
   \nonumber\\ 
& \times & \omega({\bf y''})\nabla\log\vert {\bf y''}
-({\bf x}- {\bf R}_j)\vert d^2y'd^2y''d^2x .
\label{effectiveH}
\end{eqnarray}
We can intuitively guess that it depends on the distance between center 
coordinate ${\bf R}_{ij} \equiv {\bf R}_i - {\bf R}_j$. 
The effective Lagrangian is written as 
\begin{equation} 
L_{eff} = {S_0^2\lambda \over 2}\sigma
\sum_i \left(Y_i{dX_i\over dz} - X_i{dY_i\over dz}\right) - H_{eff} .
\end{equation}
Here a special case, we consider the equation of motion for a couple of 
vortices, which becomes 
\begin{equation}
 {S_0^2\lambda \over 2}\sigma\left({\bf k}\times 
 {d{\bf R}_i \over dz}\right) 
  = - {\partial H_{eff} \over \partial {\bf R}_i} 
\end{equation}
with $i = 1, 2$. By introducing the relative coordinate 
${\bf r} \equiv {\bf R}_1 - {\bf R}_2$,we have a reduced form 
\begin{equation}
 {S_0^2\lambda \over 2}\sigma\left({\bf k}\times 
 {d{\bf r}\over dz}\right) 
  = - {\partial H_{eff} \over \partial {\bf r}} .
\end{equation}
From this we obtain the constant of motion $r^2 = C$, which is reminiscent 
of the well known conservation law of the vortex dynamics. Furthermore, we 
have the relation between the polar angle and the relative vector ${\bf r}$, 
namely, $\dot \phi = f(r) = {\rm constant}$, which shows that the two 
vortices rotate around each other by constant angular velocity 
(c.f.\cite{Rozas}), [Fig.\ref{fig2}]. Finally, we give a remark on 
a possible conversion of the effective Hamiltonian in accordance with 
the change between two types vortices (non-singular and singular types). 
Namely, for the singular vortex it is given by the logarithmic form 
$H_{eff} = -\sum_{ij}\mu_i\mu_j \log\vert {\bf R}_i - {\bf R}_j\vert$. 
On the other hand, in the limiting case that 
$\vert{\bf y'}\vert << \vert {\bf x} - {\bf R}_i \vert$, the effective 
Hamiltonian (\ref{effectiveH}) is shown to be reduced to the form 
\begin{equation}
\int \nabla \log\vert {\bf x} \vert \nabla \log\vert {\bf x}
- {\bf R}_{ij} \vert d^2x \simeq \sigma^2 
\log \vert {\bf R}_{ij} \vert 
\end{equation}
which implies that the Hamiltnian has the same form as the singular case 
except the coupling constant in such a limiting case. The difference of the 
coupling constant may play a substantial role for detecting the possible 
structure change between two types of optical vortices.

%%
%Figure caption 
%%
%%
\newpage

\begin{thebibliography}{10}
\bibitem{Chiao}
R. Y. Chiao, E. Gamire, and C. H. Townes, 
Phys. Rev. Lett. {\bf 13}, 479 (1964). 
\bibitem{Snyder}
A. W. Snyder, L. Polarian, and D.J. Mitchell, 
Opt. Lett. {\bf 17}, 789 (1992).
\bibitem{Swartz}
G. A. Swartzlander, Jr. and C. T. Law, 
Phys. Rev. Lett. {\bf 69}, 2503 (1992). 
\bibitem{Rozas}
D. Rozas, Z.S. Sacks and G.A. Swartzlander, Jr.,
Phys. Rev. Lett. {\bf 79}, 3399 (1997).
\bibitem{Born}
M. Born and E. Wolf, {\it Principle of Optics} 
(Pergamon, Oxford, 1975).
\bibitem{Landau}
L. Landau and E. Lifschitz, {\it Electrodynamics in Continuous Media}, 
Course of Theoretical Physics Vol.8 (Pergamon Oxford, 1968).
\bibitem{Kakigi}
H. Kuratsuji and S. Kakigi, Phys. Rev. Lett. 
{\bf 80}, 1888 (1998) and references cited therein. 
\bibitem{Brosseau}
C.Brosseau, {\it Fundamental of Polarized Light: A Statistical 
Optics Approach}, Chapter 3, (John Wiley, New York, 1998).
\bibitem{Ono}
H. Ono and H. Kuratsuji, Phys. Lett. {\bf 186A}, 255(1994), 
H. Kuratsuji and H. Yabu, J. Phys. A{\bf 29}, 6505(1996).
\bibitem{Note}
We note that $\sigma$ in (\ref{vorticity}) does not depend on ${\bf R}$, 
because the integrand of $\sigma$ is a function of ${\bf x} - {\bf R}$, 
and with a change of the variable ${\bf x} \rightarrow {\bf x} - {\bf R}$, 
$\sigma$ becomes independent from ${\bf R}$.
\bibitem{Lamb}
e.g. H. Lamb, {\it Hydrodymanics}, (Cambridge University Press, 1932) p248. 
\bibitem{Volovik}
M. M. Salomaa and G. E. Volovik, Rev. Mod. Phys. {\bf 59}, 533(1987).
\bibitem{potential}
Here we use the ``current function'' $\psi$ such that 
${\bf v} = {\bf k} \times \nabla\psi$, which satisfies the equation; 
$\nabla^2\psi = \omega(x)$. 
\end{thebibliography}
\end{document}